\documentstyle{mn}
\input{psfig}
\newcommand{\etal}{{et al.~}}

\newcommand{\kmsmpc}{\>{\rm km}\,{\rm s}^{-1}\,{\rm Mpc}^{-1}}

\newcommand{\cm}{\>{\rm cm}}
\newcommand{\Mpc}{\>{\rm Mpc}}
\newcommand{\kpc}{\>{\rm kpc}}
\newcommand{\Msun}{\>{\rm M_{\odot}}}

\newcommand{\keV}{\>{\rm keV}}

\def\u{{\bf u}}
\def\v{{\bf v}}
\def\w{{\bf w}}
\def\x{{\bf x}}

\def\K{{\rm \ K}}

\def\tento#1{\times 10^{#1}}

\def\gtsima{$\; \buildrel > \over \sim \;$}
\def\ltsima{$\; \buildrel < \over \sim \;$}
\def\prosima{$\; \buildrel \propto \over \sim \;$}
\def\gsim{\lower.7ex\hbox{\gtsima}}
\def\lsim{\lower.7ex\hbox{\ltsima}}
\def\simgt{\lower.7ex\hbox{\gtsima}}
\def\simlt{\lower.7ex\hbox{\ltsima}}
\def\simpr{\lower.7ex\hbox{\prosima}}

\newcommand{\apj}{ApJ}

\newcommand{\mnras}{MNRAS}

\newcommand{\nat}{Nature}

\newdimen\hssize
\newdimen\hdsize
\begin{document}


\title[The Angular Momentum of Gas in Proto-Galaxies]
      {The Angular Momentum of Gas in Proto-Galaxies: II -- 
       The Impact of Preheating}

\author[van den Bosch, Abel \& Hernquist]
       {Frank C. van den Bosch$^{1}$, Tom Abel$^{2}$, and Lars Hernquist$^{3}$
        \thanks{E-mail: vdbosch@mpa-garching.mpg.de}\\
       $^1$Max-Planck-Institut f\"ur Astrophysik, Karl Schwarzschild
           Str. 1, 85741 Garching, Germany\\
       $^2$The Pennsylvania State University, University Park, 
             PA 16802, USA \\
       $^3$Harvard Smithsonian Center for Astrophysics, Cambridge, MA 02138
           USA}


\date{}

\pagerange{\pageref{firstpage}--\pageref{lastpage}}
\pubyear{2000}

\maketitle

\label{firstpage}


\begin{abstract}
  We examine the  effect of preheating of the  intergalactic medium on
  galaxy formation using  cosmological hydrodynamical simulations.  By
  performing  simulations both  with and  without a  simple  model for
  preheating,   we   analyse   and   compare  the   angular   momentum
  distributions of the dark matter and the baryons. Preheating unbinds
  baryons  from their  dark matter  haloes, yielding  a  baryonic mass
  fraction that declines with  decreasing halo mass.  In addition, the
  spin  parameter of  the  gas is  reduced  with respect  to the  case
  without  preheating,  while  the  misalignment between  the  angular
  momentum directions  of the gas and dark  matter increases strongly.
  The angular momentum distributions  of individual haloes reveal that
  preheating  decreases   (increases)  the  mass   fraction  with  low
  (negative) specific  angular momentum.  We  discuss the implications
  of these findings for the  formation of disk galaxies in a preheated
  intergalactic medium, and compare  our results to the predictions of
  Maller \&  Dekel (2002),  who propose an  alternative interpretation
  for the origin of the angular momentum of (proto)-galaxies.
\end{abstract}




\section{Introduction}
\label{sec:intro}

Angular momentum plays a crucial role in the formation and structuring
of galaxies.  This is most  evident for disk galaxies, whose structure
and dynamics  are clearly governed by their  angular momentum content.
In current  models for  disk galaxy formation,  set forth by  White \&
Rees (1978) and Fall \& Efstathiou  (1980), it is supposed that a disk
forms out of gas that cools inside a dark matter halo while conserving
its   specific    angular   momentum   acquired    from   cosmological
torques. Under the additional assumption  that the gas and dark matter
acquire the  same quantities  of specific angular  momentum (generally
expressed  though the  dimensionless spin  parameter  $\lambda$), this
model  can  successfully explain  the  observed  distribution of  disk
scale-lengths (e.g., Mo, Mao \& White 1998; de Jong \& Lacey 2000).

Nevertheless, a proper understanding of the structure and formation of
disk galaxies  in terms of the  origin and evolution  of their angular
momentum  distributions has  proved  extremely challenging.  Foremost,
hydrodynamical  simulations  of disk  formation  that include  cooling
indicate  that,  contrary to  the  standard  assumption, the  specific
angular  momentum distribution  of  the gas  is  {\it not}  conserved.
Instead, the  gas looses a large  fraction of its  angular momentum to
the  dark  matter (Navarro  \&  Benz  1991;  Navarro \&  White  1994),
yielding disks that are an order of magnitude too small.  This problem
has  become known  as  the ``angular  momentum  catastrophe'', and  is
typically associated  with the well-known  ``over-cooling problem'' in
CDM  cosmologies  (White \&  Rees  1978;  White  \& Frenk  1991).   At
early-times gas cooling is very efficient, leading to the formation of
dense gas  clumps which  loose their orbital  angular momentum  to the
surrounding  dark  matter haloes  through  dynamical friction,  before
eventually merging to form the central disk. Therefore, some mechanism
is required to prevent or delay the cooling of the gas, so that it can
preserve  a   larger  fraction  of  its   angular  momentum.   Indeed,
simulations  in which  gas  cooling is  artificially suppressed  until
$z=1$ yield larger,  more realistic  disks (Weil,  Eke  \& Efstathiou
1998; Eke, Efstathiou \& Wright 2000).

The outstanding challenge is to identify what mechanism can accomplish
this  required  delay, and  yet  be  consistent  with observations  of
galaxies and  quasars at high  redshift.  Initial studies  focussed on
heating  from the extragalactic  ionising background.   Although these
studies   have  shown  that   an  ionising   background  can   have  a
non-negligible impact on the formation of (mainly small) galaxies, the
impact on the angular momenta of  disk galaxies was found to be either
unimportant (Vedel, Hellsten \& Sommer-Larsen 1994) or exacerbated the
problem (Navarro \& Steinmetz 1997).

More promising results have been obtained with localised feedback from
supernova explosions.   Simulations that  model this type  of feedback
typically  reveal   a  significantly  reduced,   though  not  entirely
nullified,   angular   momentum   loss  (Sommer-Larsen   \etal   1999;
Sommer-Larsen,  G\"otz \&  Portinari 2002;  Thacker \&  Couchman 2000,
2001).  However, the implications  of these results are still somewhat
unclear, in view of the difficulty of modeling the physical properties
of star-forming gas in a  cosmological context.  In their recent study
of cosmic  star formation (Springel  \& Hernquist 2003a),  Springel \&
Hernquist   (2003b)  developed   a  multiphase   description   of  the
interstellar medium and showed that  one consequence of feedback is to
pressurise the gas, altering its  equation of state at high densities.
Their  idealised  models  of  disk  formation in  dark  matter  haloes
indicate  that this process  is highly  sensitive to  the form  of the
equation of state of star-forming gas, which may help to reconcile the
earlier simulations with observations.

Furthermore,  pressure  forces  associated  with  galactic  winds  may
significantly affect the baryonic mass fractions (Scannapieco, Ferrara
\& Broadhurst  2000) and/or angular momenta (Abel,  Croft \& Hernquist
2001) of neighbouring proto-galaxies, making the impact less localised
than typically assumed.  A related  form of feedback is preheating (or
rather `reheating') of the  intergalactic medium (IGM) during an early
epoch of vigorous star  formation and/or AGN activity.  Characteristic
of  preheating   is  that  the   entropy  of  the  IGM   is  increased
substantially before the main  epoch of galaxy formation; i.e.  before
the majority  of the gas  has undergone gravitational  collapse.  This
can significantly delay  cooling (i.e., Mo \& Mao  2002) and therefore
alleviate the angular momentum catastrophe (Sommer-Larsen \etal 1999).

However,   even  if   feedback  can   prevent  the   angular  momentum
catastrophe, an  important discrepancy between the  angular momenta of
dark matter haloes and disk galaxies remains.  If the standard picture
is correct, the angular  momentum distribution of disk galaxies should
be identical  to that  of dark matter  haloes. However,  the universal
angular  momentum distribution  of dark  matter haloes  (Bullock \etal
2001),  contains  far  more  low-angular  momentum  material  than  is
typically present in disk galaxies (van den Bosch 2001; van den Bosch,
Burkert \& Swaters  2001).  In a recent paper,  Maller \& Dekel (2002;
hereafter MD02)  proposed a  toy model which  indicates that  the same
feedback  that may  solve the  angular momentum  catastrophe  can also
explain this apparent mismatch of angular momentum profiles.  Although
angular  momentum  is  commonly  thought to  arise  from  cosmological
torques  (Hoyle 1953;  Peebles 1969;  White 1984),  the MD02  model is
based on an alternate, though related picture put forward by Vitvitska
\etal  (2002) and  Maller, Dekel  \& Somerville  (2002), in  which the
final angular momentum of a dark  matter halo is simply the vector sum
of the orbital angular momenta  of all its progenitor haloes.  In this
model, most of a halo's final  angular momentum owes to the last major
merger, while  the low-angular  momentum material originates  from the
many small  satellites accreted from  largely uncorrelated directions.
If,  within this  picture, low  mass  galaxies are  largely devoid  of
baryonic  material  (because   of  feedback  processes),  the  angular
momentum of the  final disk material may have  a substantially smaller
fraction of low angular  momentum material than the corresponding dark
matter halo.  This  would explain the deficit of  low angular momentum
material in observed dwarf galaxies,  and, according to MD02, may even
result in  systems in which the  baryons have a  larger spin parameter
than the dark matter.

In an ongoing  attempt to improve our understanding  of the origin and
evolution  of the  angular momentum  of baryons,  we have  performed a
number  of  simulations  of  structure  formation  in  a  $\Lambda$CDM
cosmology {\it  without cooling}.   While unrealistic, the  absence of
cooling allows us  to better focus attention on  the impact of gravity
and (shock) heating on the angular momentum of the baryons. Therefore,
such  simulations are a  logical preliminary  step to  investigate the
build-up of angular momentum in proto-galaxies.  In the first paper in
this series (van  den Bosch \etal 2002; hereafter  Paper~I), we used a
non-radiative\footnote{We use  the term ``non-radiative',  rather than
the commonly used term ``adiabatic'', to emphasise that the simulation
does include non-adiabatic  shocks.}, hydrodynamic $N$-body simulation
to  investigate whether  the  gas and  dark  matter acquire  identical
angular  momentum   distributions.   Although  this   is  the  general
assumption, based on the fact  that the gas and dark matter experience
the same  cosmological torques, gas  and dark matter  suffer different
relaxation mechanisms  during halo collapse.  Whereas  the dark matter
undergoes collisionless virialisation  through violent relaxation, gas
achieves  hydrostatic  equilibrium by  shocking.   In principle,  this
could decouple the angular momentum  distribution of the gas from that
of  the  dark matter,  possibly  explaining  the  mismatch of  angular
momentum  profiles  discussed above  without  the  need for  feedback.
However,  we  found that  the  ``initial''  (i.e.,  prior to  cooling)
angular momentum  distributions of gas and dark  matter are remarkably
similar.   This  indicates that  shock  heating  does  not lead  to  a
decoupling of the  angular momenta of gas and dark  matter and it thus
confirms one of the two main assumptions underlying the theory of disk
galaxy formation.   However, we found  that the angular  momentum {\it
directions} of  the gas and dark  matter are poorly  aligned, and that
large  mass fractions  of  the  gas (and  dark  matter) have  negative
specific angular  momentum with respect to the  total angular momentum
vector (see also Chen \&  Jing 2002).  This indicates that disk galaxy
formation   {\it  cannot}  occur   under  detailed   angular  momentum
conservation, in  violation of  the second main  assumption underlying
the theory of disk galaxy formation.

Yoshida et al. (2003) also  report a misalignment between gas and dark
matter  total angular momenta  at high  redshifts in  simulations with
$2\times  288^3$  particles and  in  volumes  600  comoving kpc  on  a
side. The pressure forces in their calculations stem from the inherent
IGM  pressure relevant  at  the  low mass  scales  of first  structure
formation.   Their  very  high  resolution calculation  validates  our
conclusions  of Paper  I and  illustrates the  importance  of pressure
forces for the angular momentum properties of gas with respect to dark
matter.

In this paper we present a  similar simulation as in Paper~I, but this
time include a  model for preheating. We analyse  the angular momentum
distributions  of the gas  and dark  matter, which  we compare  to one
another and  to results  without preheating.  The  main goals  of this
paper are: (i) to investigate  the impact of preheating on the angular
momentum of the baryons (again in the limit without cooling), and (ii)
to test various predictions of the MD02 model.

This paper  is organised  as follows.  In  Section~\ref{sec:numsim} we
describe    the    numerical    simulations    and    the    analysis.
Section~\ref{sec:impact}  presents  the  results,  and  discusses  the
implications  for  the  formation  of  disk galaxies  in  a  preheated
intergalactic   medium.   Finally,   we  summarise   our  findings  in
Section~\ref{sec:concl}.

\section{The Impact of Preheating}
\label{sec:impact}

Preheating of  the intergalactic  medium was originally  introduced to
explain the observed X-ray  properties of clusters of galaxies (Evrard
\& Henry  1991; Kaiser 1991) and  is still considered one  of the most
likely explanations for the observed $L_X$--$T$ relation of groups and
clusters of galaxies (e.g., Balogh,  Babul \& Patton 1999; Bower \etal
2001; Tozzi \& Norman 2001; Borgani \etal 2002; Tornatore \etal 2003).
Preheating also has a number  of effects on the formation of galaxies.
Foremost, due to the  pressure increase, baryons can unbind themselves
from  their dark  matter haloes,  an effect  that is  larger  for less
massive haloes.   In addition, the hot gas  in hydrostatic equilibrium
inside  dark matter  haloes has  much shallower  density distributions
than without preheating,  which causes cooling to be  more gradual and
more simultaneous  (Mo \& Mao  2002).  This (partially)  decouples the
gas from  the hierarchical,  bottom-up structure formation,  which may
explain   why   the  colour-magnitude   relations   of  galaxies   are
``inverted''  with   respect  to  simple,   hierarchical  expectations
(Kauffmann, White \& Guiderdoni 1993;  Baugh, Cole \& Frenk 1996; Cole
\etal 2000;  van den  Bosch 2002).  A  potential problem,  however, is
that  too  much  preheating,  or  at  a too  early  stage,  can  delay
star-formation  to  unrealistically  low  redshifts  (Tornatore  \etal
2002). Preheating has also been  advocated as a means to alleviate the
X-ray halo problem (White \& Frenk 1991; Pen 1999; Benson \etal 2000).
However,  detailed  simulations by  Toft  \etal  (2002) indicate  that
simple cooling flow models  overpredict X-ray luminosities and that no
X-ray halo  problem exists.  Furthermore, the work  by Tornatore \etal
(2002) indicates that, contrary  to naive expectations, preheating can
actually {\it increase} the X-ray emissivity.

Preheating  is also  expected to  impact  on the  angular momentum  of
(proto)-galaxies.   First  of  all,  since preheating  alleviates  the
overcooling  problem, it  is  to be  expected  that it  also helps  in
reducing/solving the angular momentum catastrophe.  Second, preheating
increases  the internal  pressure of  the gas,  which causes  the {\it
baryonic} proto-galaxy during the pre-collapse era to be larger (which
tends  to  increase  the  angular  momentum because  it  enlarges  the
torquing arm)  and more  spherical (which tends  to lower  the angular
momentum because  the moment of  inertia is reduced).   Finally, after
collapse and virialisation, only a  small fraction of the baryons have
remained bound  to the  system.  Since the  outer parts  (which become
unbound) typically contain most of the high angular momentum material,
this reduction of the baryonic mass fraction is most likely associated
with a reduction of the total specific angular momentum.  Furthermore,
if  the MD02  picture  for angular  momentum  acquisition is  correct,
preheating  might  result  in  lower  mass fractions  of  low  angular
momentum material because low mass satellites will have lower baryonic
mass fractions than their more  massive counterparts. The main goal of
this paper is  to establish which of these  various effects dominates,
and to investigate  in detail what the final  outcome of preheating is
on the angular momentum content of the baryons.

Despite the large number of  studies related to preheating, the actual
energy source responsible for the heating is highly speculative.  Some
constraints  can  be  placed  however.   First of  all,  the  observed
presence of  the Ly$\alpha$ forest  indicates that preheating  can not
have  occured  uniformly  over  the  entire IGM.   This  rules  exotic
preheating-candidates,  such   as  decaying  dark   matter,  unlikely.
Furthermore, if  preheating is mainly  due to Type II  supernovae, the
IGM would  be enriched to  oxygen-abundances that are  only consistent
with observations in the central  parts of nearby galaxy clusters (see
discussion  in  Sommer-Larsen \etal  1999).   Since,  at early  times,
vigorous  star formation  (and AGN  activity) occurs  predominantly in
(proto)-groups and clusters, this may simply be another argument for a
certain  amount of  non-uniformity  in the  preheating. Despite  these
constraints,  until the  energy source  is known  and  understood, the
epoch, amount,  and (non)-uniformity of  preheating are largely  to be
considered free model parameters.

\section{Numerical Simulation}
\label{sec:numsim}

To illustrate the effects of preheating on the angular momentum of the
gas  in proto-galaxies, we  perform the  same numerical  simulation as
described  in  Paper~I,  but   modified  to  include  the  effects  of
preheating.   The  simulation computes  the  evolution  of an  initial
Gaussian  random  field of  dark  matter  particles and  non-radiative
baryons   in    a   $\Lambda$CDM   cosmology    with   $\Omega_m=0.3$,
$\Omega_\Lambda=0.7$ and  with a baryon  density of $\Omega_b  = 0.021
h^{-2}$  with $h =  H_0/(100 \kmsmpc)  = 0.67$.   We use  the smoothed
particle hydrodynamics  (SPH) code GADGET (Springel,  Yoshida \& White
2001) to evolve  the density  field inside a  box of $10  h^{-1} \Mpc$
(comoving) from  redshift $z=59$ down  to $z=3$.  The  initial density
field  is realised  adopting identical  power-spectra for  baryons and
dark  matter normalised  to $\sigma_8  =  0.9$.  We  use $128^3$  (2.1
million) gas and  dark matter particles each, with  particle masses of
$m_{\rm gas} = 6.26 \times 10^{6} h^{-1} \Msun$ and $m_{\rm DM} = 3.34
\times   10^{7}  h^{-1}   \Msun$,  respectively.    The  gravitational
softening lengths  for both the gas  and the dark matter  are $4 \kpc$
(comoving).   All  results presented  below  correspond  to the  final
output at $z=3$.

The  gas has  an initial  entropy  corresponding to  a temperature  of
$2\tento{4}\K$ for  a mean  molecular weight of  one proton  mass.  At
redshift $z=7$  we reset  the temperature of  the gas particles  to $4
\times 10^6 {\rm K}$ (again for  a mean molecular weight of one proton
mass) to mimic  the effects of preheating.  Although  this is a fairly
ad--hoc  method, especially  in terms  of the  abruptness  and spatial
uniformity of the preheating, it is adequate to illustrate the type of
changes  that  preheating  may  have  on the  final  angular  momentum
distribution of  the gas  in galaxy sized  haloes.  To place  the {\it
magnitude} of this preheating model in a broader context, we follow Mo
\&  Mao (2002),  and define  the entropy  of the  gas as  ${\cal  S} =
T/n_e^{2/3}$, with $T$ the  temperature and $n_e$ the electron density
(assuming a completely ionised gas).  For a cosmology with $\Omega_b =
0.021 h^{-2}$ and  $h=0.67$, as adopted here, the  entropy in units of
$100 \keV \cm^{-2}$ at an  overdensity $\delta$ and a redshift $z$ can
be written as
\begin{equation}
\label{entropy}
{\cal S}_{100} = \left( {T \over 4 \times 10^4 \K} \right)
(1+\delta)^{-2/3} (1+z)^{-2}
\end{equation}
(Mo \& Mao 2002). Thus, a temperature of $T=4 \times 10^6 \K$ at $z=7$
corresponds  to  ${\cal   S}_{100}=1.6  (1+\delta)^{-2/3}$,  which  is
comparable  to the  observed  entropy excess  in  clusters and  groups
(Ponman, Cannon \& Navarro 1999; Lloyd-Davies, Ponman \& Cannon 2000),
and does  not violate constraints  from the CMB (e.g.   Springel \etal
2001). Note  that the simulation  does not include  radiative cooling,
and that we  can thus not investigate whether  preheating indeed helps
to solve the angular momentum catastrophe (but see Sommer-Larsen \etal
1999).

\subsection{Halo Identification}
\label{sec:haloes}

As detailed  in Paper~I, groups  of particles are identified  with the
HOP group finder  (Eisenstein \& Hut 1998). In  each group we identify
the densest particle, and we determine the virial radius $R_{\rm vir}$
of the spherical volume, centered  on this densest particle, inside of
which the  average dark matter density is  $\Delta_{\rm vir}(z)$ times
the critical density $\rho_{\rm  crit}(z)$.  For our adopted cosmology
and at the redshift of  the output analysed $\Delta_{\rm vir} = 174.7$
(Bryan \&  Norman 1998).  Next,  we compute the center-of-mass  of all
dark matter particles inside  this spherical volume which we associate
with the  new center  of the  halo. We compute  the new  $R_{\rm vir}$
around  this  new   halo  center,  as  well  as   the  associated  new
center-of-mass.  This procedure is iterated until the distance between
the  center-of-mass and  the  adopted  halo center  is  less than  one
percent of the virial  radius $R_{\rm vir}$.  Typically, this requires
of order 2 to 5 iteration steps.

All  dark matter  and  gas particles  inside  the resulting  spherical
volume  with  radius  $R_{\rm   vir}$  are  considered  halo  members.
Although our  iterative method ensures that the  center-of-mass of the
dark matter component is similar  to the adopted halo center, the same
is not necessarily true for the gas component. We therefore remove all
haloes from our sample for  which the distance between the halo center
and  the center-of-mass  of  the  gas particles  is  larger than  $10$
percent of  $R_{\rm vir}$ (see  Paper~I for details). In  addition, if
any two haloes overlap, i.e., if the distance between the halo centers
is  less than  the sum  of  their virial  radii, we  remove the  least
massive halo from our sample.  In order to allow sufficiently accurate
measurements  of the  angular momentum  vectors  of the  gas and  dark
matter, we accept only haloes  that have more than $100$ gas particles
{\it and} more than $100$  dark matter particles.  This leaves a total
of  (only)  68 haloes.   For  comparison,  in  the simulation  without
preheating described and analysed in  Paper~I, we obtained a sample of
378  haloes (using  exactly the  same halo  identification  and sample
selection).   The reason  why so  far fewer  haloes make  it  into the
sample with preheating is that the average number of gas particles per
halo is significantly reduced (see below).

\subsection{Analysis}
\label{sec:anal}

Spin parameters  and mis-alignment angles  between gas and  dark matter
components are computed in the  standard way, as described in detail in
Paper~I.  The spin parameters of the dark matter and
the gas are defined by
\begin{equation}
\label{lambda}
\lambda_{\rm gas, DM} = \frac{|{\bf j}_{\rm gas,DM}|}{\sqrt{2} \, 
R_{\rm vir} \, V_{\rm vir}},
\end{equation}
where  ${\bf j}_{\rm  gas}$ and  ${\bf j}_{\rm  DM}$ are  the specific
angular momenta of the gas  and dark matter, respectively, and $V_{\rm
vir}  =  \sqrt{G (M_{\rm  gas}  +  M_{\rm  DM})/R_{\rm vir}}$  is  the
circular  velocity at  the  virial radius  $R_{\rm  vir}$.  The  angle
between  the  angular momentum  vectors  of  both  mass components  is
defined as
\begin{equation}
\label{angle}
\theta = \cos^{-1} \left[ \frac{{\bf J}_{\rm gas} \cdot {\bf J}_{\rm DM}} 
{|{\bf J}_{\rm gas}|\, |{\bf J}_{\rm DM}|} \right],
\end{equation}
with  ${\bf J}_{\rm  gas}$ and  ${\bf J}_{\rm  DM}$ being the  total 
angular momentum vectors of the gas and dark matter, respectively.

Following Paper~I, for haloes with more than 5000 particles (in total)
we compute  the detailed distributions of specific  angular momenta of
the gas and DM particles. Both the dark matter and the gas are fluids,
for  which the  velocity of  each  {\it microscopic}  particle can  be
written as $\v = \u + \w$.   Here $\u$ is the mean streaming motion at
the  location  $\x$ of  the  microscopic  particle,  and $\w$  is  the
particle's random motion.   The velocities of the DM  particles in the
simulation correspond to $\v$ and are obtained from Newton's equations
of motion.  In the case of the gas, however, the SPH approach is used,
which  yields the  streaming  motions  $\u$ of  the  gas particles  by
solving the  Euler equations (with  some artificial viscosity  to take
account of  shocks).  Information about  the random motion of  the gas
particles is provided by the internal energy of each particle.

Thus, the velocities of the dark matter particles and gas particles in
the simulation correspond to different motions. To compare the angular
momentum distributions of the gas and dark matter in a meaningful way,
however, the same velocities have to be used. Ideally, one would focus
on the  angular momentum distributions based on  the streaming motions
$\u(\x)$.  After  all, when the  gas cools $\v \rightarrow  \u$, i.e.,
the typical  random motions  become negligible.  Under  the assumption
that the  gas conserves its  specific angular momentum,  the resulting
disk  will  thus have  an  angular  momentum  distribution related  to
$\u$. However,  this requires estimating the streaming  motions of the
dark matter. In principle this could be achieved by smoothing the dark
matter velocities $\v$, but it  is unclear what smoothing scale to use
and  resolution issues  are  likely  to play  an  important role.   We
therefore do not attempt to compute the dark matter streaming motions,
but  instead  use the  velocities  $\v$ of  the  dark  matter and  gas
particles  when comparing  their angular  momentum  distributions.  To
compute the microscopic  velocities $\v$ of the gas  particles, we use
the same `thermal broadening'  technique as described in Paper~I.  For
each gas particle we draw $100$ random velocities $\w$ which we add to
the particle's streaming motion $\u$.  For each of the three Cartesian
components of $\w$ we draw a  velocity from a Gaussian with a standard
deviation given by
\begin{equation}
\label{veldisp}
\sigma = \sqrt{k \, T \over \mu} = \sqrt{2 \, U \over 3}
\end{equation}
Here $U$ and $T$ are the internal energy per unit mass and temperature
of  the gas particle,  respectively, $k$ is  Boltzmann's constant, and
$\mu$ is the mean molecular weight of the  gas. For each particle (gas
and dark  matter)   we compute  the component  $j^v$   of the specific
angular momentum ${\bf j}^v = {\bf r} \times {\bf  v}$ that is aligned
with the {\it  total} angular momentum  vector ${\bf J}$.  We  use the
superscript $v$ to  distinguish it from  the specific angular momentum
${\bf j} = {\bf r}  \times {\bf u}$,  which is only accessible for the
baryons.
\begin{figure}
\centerline{\psfig{figure=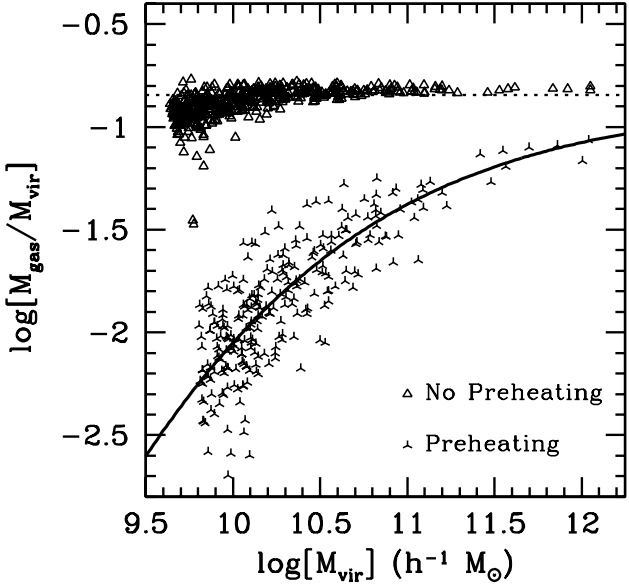,width=\hssize}}
\caption{Baryonic  mass  fractions,   $M_{\rm  gas}/M_{\rm  vir}$,  as
function of  halo virial  mass, $M_{\rm vir}$,  for the haloes  in the
simulations  without (triangles) and  with (tripods)  preheating.  The
horizontal, dotted  line corresponds to the  universal baryon fraction
$f_{\rm bar} = \Omega_b  / \Omega_m$.  In the no-preheating simulation
the baryonic  mass fractions  of all haloes  are very similar  to this
universal value.  In the  preheating case, however, gas mass fractions
are  systematically lower,  an effect  that is  larger for  lower mass
haloes.   The  thick  solid  line  is the  best-fit  relation  of  the
form~(\ref{preheat}).}
\label{fig:barfrac}
\end{figure} 
\begin{figure*}
\centerline{\psfig{figure=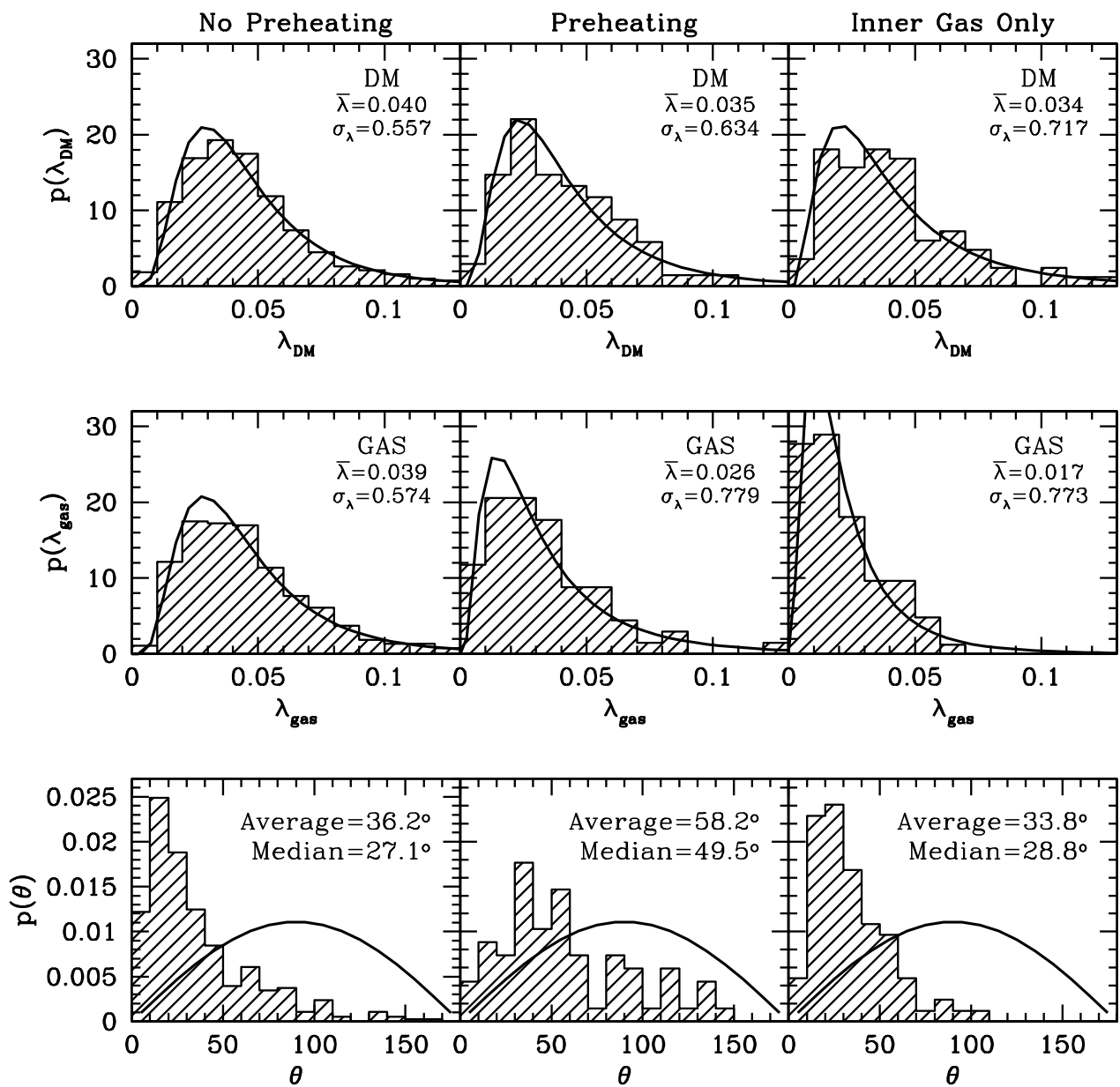,width=0.95\hdsize}}
\caption{Distributions   of   $\lambda_{\rm   DM}$   (upper   panels),
$\lambda_{\rm  gas}$  (middle  panels),  and  the  misalignment  angle
between the  angular momentum vector of  the gas and  the dark matter,
$\theta$  (lower  panels).   Results  are  shown  for  haloes  in  the
simulation  without  (left-hand   panels)  and  with  (middle  column)
preheating.   In addition,  we show  the results  for the  ``inner gas
only''  sample  of  haloes  (right-hand  panels)  extracted  from  the
no-preheating simulation as described in  the text. The solid lines in
the  upper and  middle panels  correspond to  the  best-fit log-normal
distributions  (eq.~[\ref{spindistr}]), with  the  best-fit values  of
$\bar{\lambda}$ and $\sigma_{\lambda}$ as indicated. The solid line in
the lower panels correspond to a random distribution of angles, and is
shown  for comparison.   See text  for  a detailed  discussion of  the
results.}
\label{fig:lambdas}
\end{figure*}

\section{Results}
\label{sec:results}

\subsection{Baryonic mass fractions}
\label{sec:barfrac}

Fig.~\ref{fig:barfrac}  plots  the   baryonic  mass  fraction  $M_{\rm
gas}/M_{\rm vir}$  as a  function of halo  virial mass $M_{\rm  vir} =
M_{\rm  gas} +  M_{\rm  DM}$.  Results  are  shown for  the case  with
(tripods)  and without (triangles)  preheating.  Whereas  the baryonic
mass fractions in haloes without preheating are close to the universal
value $f_{\rm  bar} = \Omega_b  / \Omega_m$ (indicated  by horizontal,
dotted line),  in the  preheating simulation the  gas mass  within the
virial radii  of dark matter haloes is  significantly suppressed. This
suppression is larger for smaller mass haloes and can be parameterised
using the fitting formula suggested by Gnedin (2000):
\begin{equation}
\label{preheat}
M_{\rm gas} = {f_{\rm bar} M_{\rm DM} \over \left[1 + (2^{\alpha/3} - 1)
\left({M_f \over M_{\rm DM}}\right)^{\alpha} \right]^{3/\alpha}}
\end{equation}
Here $\alpha$ and  the filter mass $M_f$ are  free fitting parameters.
Our   best-fit  relation   is  shown   as  a   thick  solid   line  in
Fig.~\ref{fig:barfrac}  and  has  $\alpha=0.4$  and $M_f  =  5  \times
10^{11} h^{-1} \Msun$.
\begin{figure*}
\centerline{\psfig{figure=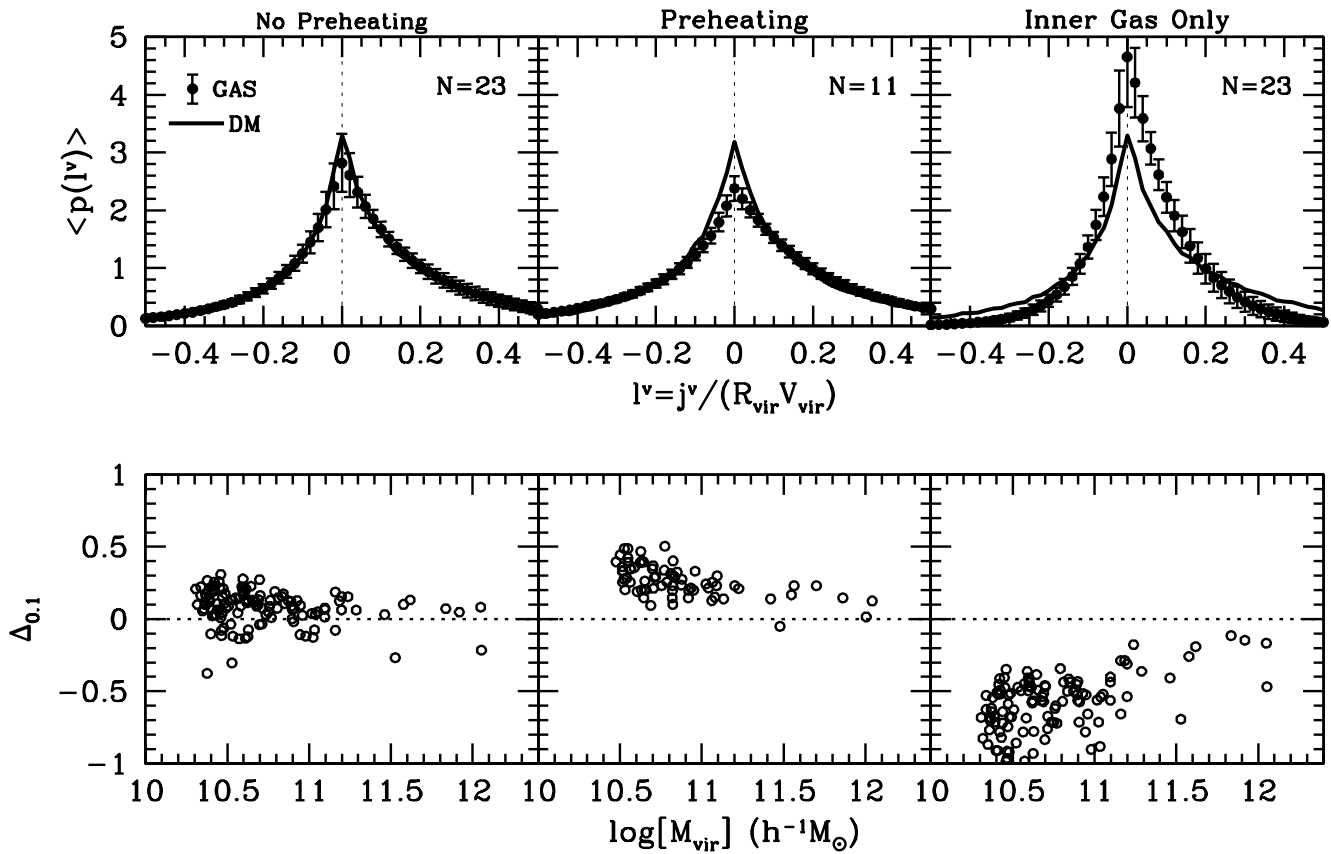,width=\hdsize}}
\caption{Upper  panels plot the  distribution of  normalised, specific
angular momentum, $l^v$, averaged  over all haloes (the actual number,
$N$,  is indicated in  each panel)  with more  than 5000  particles in
total. Results  are shown  for three samples  of haloes,  as indicated
above each  panel. Solid lines and dots  with errorbars (corresponding
to the  rms scatter) correspond to the  angular momentum distributions
of the dark matter and  gas, respectively. Whereas the $\langle p(l^v)
\rangle$ of the gas and  dark matter are indistinguishable in the case
without preheating,  the gas lacks predominantly  low angular momentum
material  with respect  to  the  dark matter  in  the simulation  with
preheating. The  opposite is true for  the inner gas  mass fraction in
haloes  without  preheating  (right-hand  panel).  These  results  are
confirmed   in   the   lower   panels,   which   plot   $\Delta_{0.1}$
(equation~[\ref{deltastat}])  as function  of halo  virial  mass.  The
quantity $\Delta_{0.1}$  is a measure  for the difference  between the
mass fractions of dark matter and gas with $\vert l^v \vert < 0.1$. In
the  no-preheating simulation,  $\Delta_{0.1} \sim  0$  independent of
halo  mass.  With preheating,  however, $\Delta_{0.1}$  decreases with
decreasing  halo mass, consistent  with the  predictions of  Maller \&
Dekel  (2002). The  ``inner  gas only''  sample  reveals the  opposite
behaviour,  indicating  that  low  angular momentum  material  resides
predominantly near the center of the potential well.}
\label{fig:lowl}
\end{figure*}

We  can  use  this  fitting  function  to gain  some  insight  in  how
preheating affects  the angular momentum distribution  of the baryons.
We reanalyse  the no-preheating  simulation presented in  Paper~I, but
this time considering only the inner-most gas mass fraction whose mass
is  given  by equation~(\ref{preheat})  with  $\alpha=0.4$ and  $M_f=5
\times  10^{11}  h^{-1}  \Msun$.   Applying the  same  halo  selection
criteria as described in  Section~\ref{sec:haloes} we end up with $83$
haloes, which we analyse exactly as before.  In what follows, we refer
to this  sample of haloes  as the ``inner  gas only'' sample.   If the
differences between  the no-preheating and  preheating simulations are
predominantly due to  the fact that in the latter  only the most bound
fraction of the gas remains within the virial radius, this ``inner gas
only'' sample should reveal  similar angular momentum distributions as
in the preheating simulation.

\subsection{Spin parameters \& misalignment angles}
\label{sec:spin}

In Fig.~\ref{fig:lambdas}, we compare the spin parameter distributions
of  the gas and  dark matter,  and contrast  the simulations  with and
without preheating.  The hatched  histograms in the upper, middle, and
lower panels plot the distributions  of the spin parameter of the dark
matter,  $p(\lambda_{\rm DM})$,  of  the spin  parameter  of the  gas,
$p(\lambda_{\rm gas})$, and of  the angle between the angular momentum
vectors of both mass  components, $p(\theta)$, respectively. Panels on
the  left  correspond  to  the  simulation  without  preheating  (cf.,
Paper~I),  panels in  the  middle correspond  to  the simulation  with
preheating,  and panels  on  the right  correspond  to the  simulation
without preheating, but in which only the inner fraction of the gas is
used in the  analysis (as described above).  The  thick solid lines in
the  upper   and  middle  panels  indicate   the  best-fit  log-normal
distributions
\begin{equation}
\label{spindistr}
p(\lambda){\rm d} \lambda = {1 \over \sqrt{2 \pi} \sigma_{\lambda}}
\exp\biggl(- {{\rm ln}^2(\lambda/\bar{\lambda}) \over 2
  \sigma^2_{\lambda}}\biggr) {{\rm d} \lambda \over \lambda}.
\end{equation}
with  the best-fit  values of  $\bar{\lambda}$  and $\sigma_{\lambda}$
indicated. The thick  solid lines in the lower  panels correspond to a
random distribution of angles $\theta$, and are shown for comparison.

As already  discussed in Paper~I, the spin  parameter distributions of
the gas  and dark  matter are remarkably  similar in the  case without
preheating.   However,  in the  preheating  simulation presented  here,
there  is a clear  indication that  $\lambda_{\rm gas}  < \lambda_{\rm
DM}$ on average. For the  dark matter $\bar{\lambda}_{\rm DM} = 0.035$
and  $\sigma_{\lambda_{\rm  DM}}  =  0.63$ (almost  identical  to  the
no-preheating  case), while  for  the gas  $\bar{\lambda}_{\rm gas}  =
0.025$  and  $\sigma_{\lambda_{\rm  gas}}  =  0.76$.   Preheating  has
reduced the typical specific angular  momentum of the gas with respect
to that of the dark matter and  with respect to that of the gas in the
no--preheating  case.  In  addition,   with  an  average  (median)  of
$58.2^{\rm o}$ ($49.5^{\rm o}$)  the $\theta$-distribution in the case
with preheating is clearly different from the case without preheating.
Although it  is still  significantly different from  that of  a purely
random  distribution,  preheating has  largely  decoupled the  angular
momentum directions of the gas and dark matter.

If   the  differences   between  the   no-preheating   and  preheating
simulations result predominantly from the fact that in the latter case
only the  most bound  fraction of  the gas ends  up inside  the virial
radius,  the haloes  of the  ``inner gas  only'' sample  should reveal
similar distributions of $\lambda_{\rm gas}$ and $\theta$ as for those
with   preheating.   This,   however,   is  clearly   not  the   case:
$p(\lambda_{\rm  gas})$ is  shifted to  even lower  values,  while the
distribution of  $\theta$ is virtually indistinguishable  from that of
the   no-preheating  simulation  (and   thus  differs   strongly  when
preheating is  included).  The increase  in the misalignment  found in
the preheating simulation is thus not merely due to the fact that only
the inner fraction  of the gas remains bound  to the halo.  Apparently
preheating has modified the inertia  tensor of the gas with respect to
that  of the  dark  matter, which  has  resulted in  a larger  average
misalignment  between  the two  mass  components.   In addition,  this
comparison indicates that the fraction  that remains bound in the case
with preheating  has acquired {\it more} specific  angular momentum as
the same mass in the case without preheating.  This is due to the fact
that the extent  (and thus the torquing arm) of that  gas is larger in
the  case  of  preheating,  and/or,  to differences  in  the  relative
importance of low mass progenitors, as suggested by MD02.

\subsection{Angular Momentum Distributions}
\label{sec:distr}

In the previous section we  compared the distributions of the (global)
spin  parameters  of  dark  matter  and gas,  both  with  and  without
preheating.   Here,  we  compare  the angular  momentum  distributions
within individual haloes.

As described  in Section~\ref{sec:anal}, for each  particle we compute
the specific angular  momentum $j^v$ along the direction  of the total
angular momentum  vector. Following  Paper~I we define  the normalised
specific angular momenta  $l^v = j^v / (R_{\rm  vir} V_{\rm vir})$ and
compute separate,  normalised distribution functions  $p(l^v)$ for the
gas  and  dark  matter  in  each individual  halo.   Because  of  this
normalisation, the  distributions of different haloes  can be compared
directly,  and  be used  to  compute  the  {\it average}  distribution
function
\begin{equation}
\label{averdf}
\langle p(l^v) \rangle = {1 \over N} \sum_{i=1}^{N} p_i(l^v)
\end{equation}
where the average  is taken over the $N$ haloes  with more than $5000$
particles in  total.  In the  simulation without preheating  there are
$N=23$ haloes that  make our selection criterion, which  is reduced to
$N=11$   in  the   case  with   preheating.   The   upper   panels  of
Fig.~\ref{fig:lowl} plot  $\langle p(l^v)  \rangle$ for both  the dark
matter  (solid  line)  and   the  gas  (solid  dots).   The  errorbars
correspond to the rms of the scatter about the mean, and, for clarity,
are only plotted for the gas.

The upper left panel  of Fig.~\ref{fig:lowl}, which corresponds to the
simulation  without preheating, confirms  the conclusion  from Paper~I
that the angular momentum distributions of the gas and dark matter are
virtually  identical,  and thus  that  the  shocks  that occur  during
virialisation  do  not   significantly  modify  the  angular  momentum
distribution of the gas.  The upper middle panel, which corresponds to
the simulation with preheating, however,  shows a clear deficit of low
angular momentum material for the gas with respect to that of the dark
matter.  This is  in good  agreement  with predictions  from the  MD02
model, even though the effect shown is fairly small. However, this may
be due to the fact that we have included only massive haloes with more
than 5000 particles, while MD02 predict that the effect is stronger in
less massive haloes.  In order to investigate this mass dependence and
to  make the  differences between  $p(l^v_{\rm gas})$  and $p(l^v_{\rm
DM})$ at low $l^v$ more quantitative, we define the statistic
\begin{equation}
\label{deltastat}
\Delta_{0.1} = 1 - {
\int_{-0.1}^{0.1} p(l^v_{\rm gas}) \, {\rm d}l^v_{\rm gas} \over 
\int_{-0.1}^{0.1} p(l^v_{\rm DM})  \, {\rm d}l^v_{\rm DM}}
\end{equation}
which  is a  measure of  the difference  between the  dark  matter and
baryonic mass  fractions with  $\vert l^v \vert  \leq 0.1$.  The lower
panels  of  Fig.~\ref{fig:lowl}  plot  $\Delta_{0.1}$ as  function  of
virial  mass.  In  order to  increase  the number  statistics we  have
included all  haloes with more  than 1000 particles in  total. Whereas
$\Delta_{0.1} \sim 0$,  independent of halo mass, in  the case without
preheating, there  is a clear trend of  decreasing $\Delta_{0.1}$ with
decreasing halo mass in the case with preheating.  This indicates that
lower mass  haloes contain relatively smaller  baryonic mass fractions
with low angular momentum, in excellent agreement with the MD02 model.
\begin{figure}
\centerline{\psfig{figure=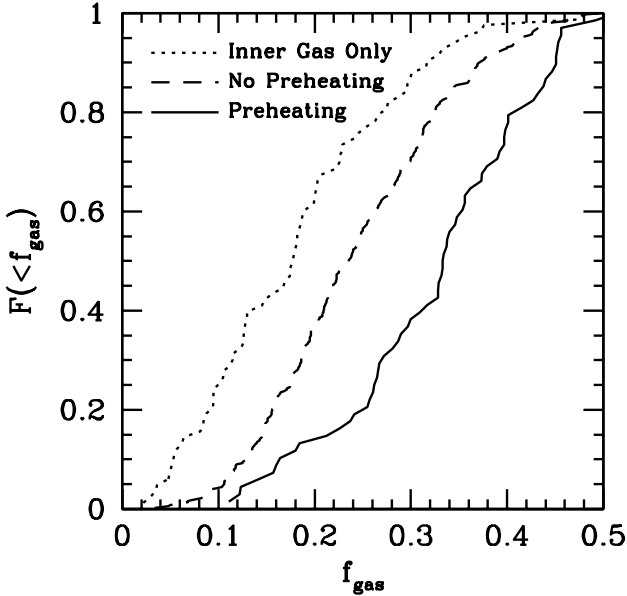,width=\hssize}}
\caption{Cumulative distributions of  the baryonic mass fractions with
$j<0$ (with $j$  determined from the actual {\it  streaming} motion of
the  gas).  Results  are shown  for the  ``preheating''  ($N=68$), the
``no-preheating''  ($N=378$)  and  the  ``inner  gas  only''  ($N=83$)
samples,  as indicated.   Simple  K-S tests  indicate  that all  three
distributions are very significantly different from one-another.  Note
that  the ``preheating'' case  reveals the  distribution that  is most
skewed towards relatively large  mass fractions with negative specific
angular  momentum.   Apparently,  preheating  causes  an  increase  in
$f_{\rm gas}$ and thus in  the decoherence of the streaming motions of
the gas.}
\label{fig:cuml}
\end{figure}

The  right-hand  panels of  Fig.~\ref{fig:lowl}  show $\langle  p(l^v)
\rangle$ and  $\Delta_{0.1}(M_{\rm vir})$  for the ``inner  gas only''
sample. Here,  the baryonic material  clearly has much {\it  more} low
angular  momentum material than  the dark  matter.  In  addition, this
effect becomes stronger  for lower mass haloes. A  comparison with the
preheating  results  clearly proves  that  preheating  has not  merely
``stripped'' the outer layers of  gas from the dark matter haloes, but
has  also affected  the actual  angular momentum  distribution  of the
remaining gas.

In addition  to the  mass fraction of  low angular  momentum material,
another useful statistic  regarding the angular momentum distributions
are the  fractions $f^v_{\rm DM}$  and $f^v_{\rm gas}$ of  dark matter
and  gas particles,  respectively, that  have {\it  negative} specific
angular momentum. Note that negative  here means that the angle $\phi$
between the  total angular momentum vector and  the particle's angular
momentum vector ${\bf j}^v = {\bf r} \times {\bf v}$ lies in the range
$90^{\rm o} <  \phi \leq 180^{\rm o}$.  As shown  in Paper~I, $f^v$ is
anti-correlated with the spin parameter, such that systems with higher
specific angular  momentum have lower  mass fractions with $j^v  < 0$.
As can be seen from the average distributions $\langle p(l^v) \rangle$
shown  in the  upper panels  of Fig.~\ref{fig:lowl}  the  average mass
fraction with  $j^v < 0$ is  fairly large ($\sim  40$ percent).  This,
however, is to a large extent a reflection of the fact that the random
motions $\w$ are larger than the streaming motions $\u$.  As indicated
in  Section~\ref{sec:anal} (see also  Paper~I), the  more illustrative
angular  momentum  distribution is  the  one  based  on the  streaming
motions, $p(l)$.   Here $l=j/(R_{\rm vir}  V_{\rm vir})$ with  $j$ the
component of the  specific angular momentum vector ${\bf  j} = {\bf r}
\times  \u$ that  is aligned  with the  total angular  momentum vector
${\bf J}$. As  shown in Paper~I, $p(l)$ typically  has a less extended
wing to  negative specific angular  momentum than $p(l^v)$ and  a more
pronounced peak at low angular momentum.

Fig.~\ref{fig:cuml}  plots the {\it  cumulative} distributions  of the
baryonic mass  fractions, $f_{\rm  gas}$, with $l  < 0$.   Results are
shown for all three samples and take all haloes with more than 100 gas
particles   and   more   than   100   dark   matter   particles   into
account. Although the median of these distributions is clearly smaller
than the  $\sim 0.4$ in the  case of $f^v_{\rm gas}$,  they indicate a
significant amount of decoherence in the streaming motions of the gas.
As detailed in Paper~I, this  implies that disk formation cannot occur
under detailed  conservation of specific angular  momentum, posing new
challenges to the usual picture of disk formation.  In the ``inner gas
only'' case, the distribution of  $f_{\rm gas}$ is shifted slightly to
lower  values  compared  to  the  no-preheating case.   The  K-S  test
indicates a probability of $10^{-5}$ that both are drawn from the same
distribution.   This  indicates  that  the amount  of  decoherence  is
radially  dependent,  in  agreement  with  the  notion  that  material
accreted  at different  times introduces  a radial  dependence  of the
angular momentum direction (Ryden 1988; Quinn \& Binney 1992).  In the
case {\it  with} preheating,  the median of  $p(f_{\rm gas})$  is {\it
larger} than  in the  case without preheating.   According to  the K-S
test, the  probability that both  $p(f_{\rm gas})$ are drawn  from the
same parent distribution is  only $9.8 \times 10^{-9}$. This indicates
that  preheating  significantly   increases  the  mass  fraction  with
negative specific angular momentum.

\section{Discussion \& Conclusions}
\label{sec:concl}

A full understanding  of the structure and formation  of disk galaxies
within a  hierarchical, cold dark  matter cosmogony faces a  number of
intriguing challenges.  First, in  the absence of any heating, baryons
cool extremely efficiently, producing  too many satellite galaxies and
resulting  in   what  has  become   known  as  the   angular  momentum
catastrophe. In  addition, the angular momentum  distributions of both
the gas and the dark  matter in numerical simulations reveal an excess
of  low angular  momentum  material compared  to  real disk  galaxies.
Moreover, as shown  by van den Bosch \etal  (2002), virialised systems
reveal  large  mass fractions  with  {\it  negative} specific  angular
momentum.  Since disks typically  contain no negative specific angular
momentum  material,   disk  formation  cannot   occur  under  detailed
conservation  of  specific  angular  momentum,  contrary  to  standard
assumptions.

It  is  generally hoped  that  these  problems  subside when  feedback
related  heating  effects are  included.  Arguably,  the most  popular
feedback effect is  heating of the cold gas in  galaxies by the energy
feedback from  star formation.   In semi-analytical models  for galaxy
formation this type of feedback is typically implemented in such a way
that  large fractions  of the  cold gas  are ejected  from  the galaxy
and/or  the dark matter  halo. Although  these ``galactic  winds'' may
successfully explain  several of the observed  properties of galaxies,
it remains  unclear whether this picture is  realistic.  For instance,
detailed hydrodynamical  simulations have  shown that the  actual mass
ejection efficiencies may be  much smaller than typically assumed (Mac
Low \& Ferrara 1999; Strickland \& Stevens 2000).

The preheating  investigated in this  paper is an alternative  form of
feedback which relies  on heating the gas before it  becomes part of a
virialised dark  matter halo.  This  type of feedback has  mainly been
discussed in connection with the observed entropy excess in groups and
clusters of  galaxies (e.g., Kaiser  1991; Ponman \etal  1999; Borgani
\etal 2002), but might be of equal importance for galaxy formation (Mo
\&  Mao   2002).   However,  unlike   the  supernova-induced  feedback
mentioned  above,   the  energy   source  for  preheating   is  highly
speculative.   Possible candidates  include  high-redshift (starburst)
galaxies, population  III stars, active  galactic nuclei, or  even the
decay of some dark matter  species. Therefore, until the energy source
is known and understood,  the epoch and (non)-uniformity of preheating
are    to     be    considered    free     model    parameters    (see
Section~\ref{sec:impact} for a more detailed discussion).

In this paper we investigated  the impact of preheating on the angular
momentum of gas  in proto-galaxies.  To that end,  we used a numerical
simulation of structure formation in a standard $\Lambda$CDM cosmology
with  non-radiative  gas,  in  which  we  mimic  an  extreme  form  of
preheating  by   instantaneously  and  homogeneously   increasing  the
temperature of  the gas at $z=7$.  Although  somewhat unrealistic, the
amount  of  entropy injected  into  the  gas  is consistent  with  the
observed  entropy  excess  in  groups  and clusters  of  galaxies.  By
comparing the  angular momentum distributions  of gas and  dark matter
both  in simulations  with and  without preheating  we  identified the
following effects;

\begin{itemize}

\item preheating unbinds baryons from dark matter haloes, resulting in
a baryonic mass fraction that decreases with decreasing halo mass.

\item with preheating, the spin  parameter of the gas within the virial
radius is smaller than that of all gas within the virial radius in the
case without  preheating, but larger than  that of the  inner most gas
with the same mass.

\item preheating largely decouples  the angular momentum directions of
the gas and dark matter.

\item  preheating  decreases  the  baryonic mass  fractions  with  low
specific angular momentum.

\item preheating  increases the baryonic mass  fractions with negative
specific angular momentum.

\end{itemize}

These  effects   are  most  easily  understood   by  considering  that
preheating increases the internal pressure of the gas. This means that
the gas associated with a proto-galaxy becomes more extended, and more
spherical in shape. Part of the gas can become unbound, which explains
the  reduced baryonic  mass  fractions.  In  addition, this  partially
explains why the  spin parameter of the gas that  remains bound to the
dark matter haloes  is reduced.  The baryons that  have become unbound
were  mostly located  at larger  radii, which  experience  the largest
torquing forces.  Thus, the  unbinding preferentially removes the high
angular momentum  material. In addition, the internal  pressure of the
gas  modifies  the moment  of  inertia with  respect  to  that of  the
underlying  dark matter,  which  explains why  the  directions of  the
angular  momentum  vectors of  the  gas and  the  dark  matter are  so
strongly  decoupled.  The  reduction  in the  mass  fraction with  low
angular momentum may  be understood as due to  the modification of the
density profile of the gas due to the increased internal pressure.  As
shown  by  Mo \&  Mao  (2002),  preheating  creates extremely  shallow
density  distributions  for  the   gas.   This  means  that  there  is
relatively less  material with a small  moment of inertia,  and thus a
reduction in the fraction of low angular momentum material.

The  above interpretation  is  based  on the  standard  idea that  the
angular  momentum  results  from  cosmological  torques  acting  on  a
proto-galaxies with a non-zero  moment of inertia.  However, in recent
papers Vitvitska \etal (2002) and Maller \etal (2002) have proposed an
alternative model, in which the angular momentum is due to the orbital
angular momentum  from the progenitors.   Maller \& Dekel  (2002) have
presented a toy-model based on  this picture that includes the effects
of  feedback. They  argue that  as long  as feedback  effects  imply a
baryonic mass fraction that declines  with decreasing halo mass (as is
the  case for the  preheating model  presented here),  the gas  in the
final halo should have: (i)  a spin parameter that is typically larger
than  that of  its  dark matter  halo,  and (ii)  an angular  momentum
distribution with  relatively less low angular  momentum material. Our
results disagree with (i),  but confirm (ii), including the prediction
that the effect  should be stronger for less  massive haloes. At first
sight it may  seem contradictory that we find a  reduction of the mass
fraction  with low  angular  momentum,  yet a  decrease  in the  total
angular momentum. The explanation,  however, is that the mass fraction
with  negative specific angular  momentum has  increased, which  has a
stronger  impact on  $\lambda_{\rm  gas}$ than  the  reduction of  low
angular momentum material.

The  origin  of this  increase  in  the  mass fraction  with  negative
specific angular momentum  is not clear, neither in  the MD02 scenario
nor  in  the cosmological  torque  picture.   In  principle, the  {\it
presence}  of  negative  specific  angular  momentum  is  more  easily
understood in terms of  the satellite accretion model.  However, there
is  no obvious  explanation for  why the  amount of  negative specific
angular momentum  should increase  in a preheated  IGM.  Note  that we
have focused  on the  simplest possible form  of the role  of pressure
forces namely  uniform preheating  at high redshifts.   Our particular
finding  that preheating  increases the  baryonic mass  fractions with
negative  specific  angular  momentum  may  change  as  one  considers
different forms of feedback.   For example, winds produced by galaxies
with  high star  formation rates  are likely  to produce  complex flow
patterns around  them, as  illustrated by figures  5-8 in  Springel \&
Hernquist (2003b).   The pressure forces arising from  such winds will
act  differently  than  the  ones  caused by  our  preheating  scheme.
Studying these more subtle  differences in feedback prescriptions will
require a  large suit of higher resolution  numerical simulations that
are beyond the scope of the present investigation.

Our  results have  important implications  for the  formation  of disk
galaxies. As we have shown,  preheating can regulate the baryonic mass
fractions  of dark  matter haloes  in much  the same  way  as galactic
winds, and should therefore be  able to alleviate the angular momentum
catastrophe.  Indeed,  simulations by Sommer-Larsen  \etal (1999) that
include both cooling and preheating find a reduction (although modest)
of  the  angular  momentum  loss  of the  baryons.   Furthermore,  the
reduction  of the  baryonic mass  fraction with  low  angular momentum
material  is  advantageous  for  the  formation  of  realistic  disks.
However, preheating also has some effects that are less favourable for
disk formation. First  of all, the total specific  angular momentum of
the gas within  the virial radius, and which is  thus eligible to cool
and form  a disk galaxy, is reduced  with respect to that  of the dark
matter.   Thus, although  the angular  momentum loss  may  be reduced,
there is  less angular momentum  to start with.  Second,  the detailed
angular momentum distributions reveal a clear increase of the baryonic
mass  fractions with  negative specific  angular momentum,  making the
formation  of a disk  dominated galaxy  less plausible.   We therefore
conclude  that  understanding  disk  formation remains  an  intriguing
puzzle, even in a preheated IGM.


\section*{Acknowledgements}

We thank  Houjun Mo and  Simon White for stimulating  discussions, and
the  referee,  Jesper  Sommer-Larsen,  for his  comments  that  helped
improve the  presentation of  the paper.  This  work was  supported in
part through  NSF grants  ACI96-19019, AST-9803137, AST  99-00877, and
AST 00-71019.  and through the Grand Challenge Cosmology Consortium.


\label{lastpage}

\end{document}